\documentclass[11pt]{article}
\usepackage[a4paper,margin=2.5cm]{geometry}
\usepackage{amsmath,amssymb,bm}
\usepackage{microtype}
\usepackage{graphicx}
\usepackage{color}
\title{After thermalization}
\author{Jorge Kurchan~ \\~\\
Laboratoire de Physique de l'\'{E}cole Normale Sup\'erieure, ENS, Universit\'e PSL, CNRS,\\
Sorbonne Universit\'e, Universit\'e de Paris, F-75005 Paris, France}
\date{}

\begin{document}
\maketitle
\begin{abstract}
  
`Complexity' measures the exploration of state space by the dynamics at a finer level of detail than that required for the equilibration of standard extensive observables \cite{Susskind2020ThreeLectures}, and therefore continues to grow long after these observables have equilibrated. In a recent paper \cite{CaoKurchan2026}, we proposed that this exploration may be quantified by  the large deviations attained over time, or, equivalently, by the time taken for `rare' observables -- such as, for example the exponential of the magnetization --    to equilibrate.
 It turns out that the evolution of complexity can then be naturally mapped onto a standard model of glassy dynamics, the "Trap Model" \cite{Bouchaud1992}, which exhibits aging and effective temperatures. Under this mapping, the exploration of increasingly fine details of phase-space corresponds to the discovery of progressively deeper minima in the glassy model.
\end{abstract}

\section{Introduction}
After a certain time, a system  thermalizes: all extensive quantities constructed from local observables fluctuate around their canonical or microcanonical averages. This does not mean that the space has been  explored  in great detail, but rather that its sampling is good enough for the purposes of `reasonable' (extensive, local) operators. Hence, equilibration times are often quite short.
Does something keep happening afterwards? A quick answer is {\it `yes, but'}. For example, once the magnetization $M$ settles around its equilibrium value, a quantity like $e^{aM}$ continues to evolve long afterward, {\it and has its own well-defined `equilibration' (exponential) time}. But why would one be interested in knowing this?
A  motivation \cite{Susskind2020ThreeLectures} comes from the duality between gravity and quantum field theory. The equilibration of a black hole's horizon is fast, but its interior keeps evolving for much longer times, raising the question of what the counterpart of this post-equilibration evolution is on the quantum field theory side.

In Susskind's proposal, quantum complexity is the minimum number of elementary operations needed to approximate the   the time-evolved state or the unitary operator to precision $\epsilon$. A related idea due to Nielsen~\cite{nielsen06-1,nielsen06} is to
 define quantum  complexity as the length of a shortest geodesic on the unitary group, equipped with an appropriate  cost function.
It is expected to grow until $
t_{\mathrm{plateau}}\sim e^N\log\left(\frac{1}{\epsilon}\right)$
and to 
reach a plateau
$\mathcal C_{\max}\sim e^N\log\left(\frac{1}{\epsilon}\right)
$.

Another much studied alternative is Krylov complexity~\cite{uogh,Rabinovici:2020ryf,vijay,nandy-review,Rabinovici:2025otw}, which measures the spreading of a wavefunction (or operator) in the Krylov basis, constructed  via de Lanczos iteration. This spreading typically continues until $t\sim e^N$.
The two notions are to some extent related~\cite{krylov-nielsen}.

\section{Complexity as large deviations}

\begin{figure}
    \centering
    \includegraphics[width=1\linewidth]{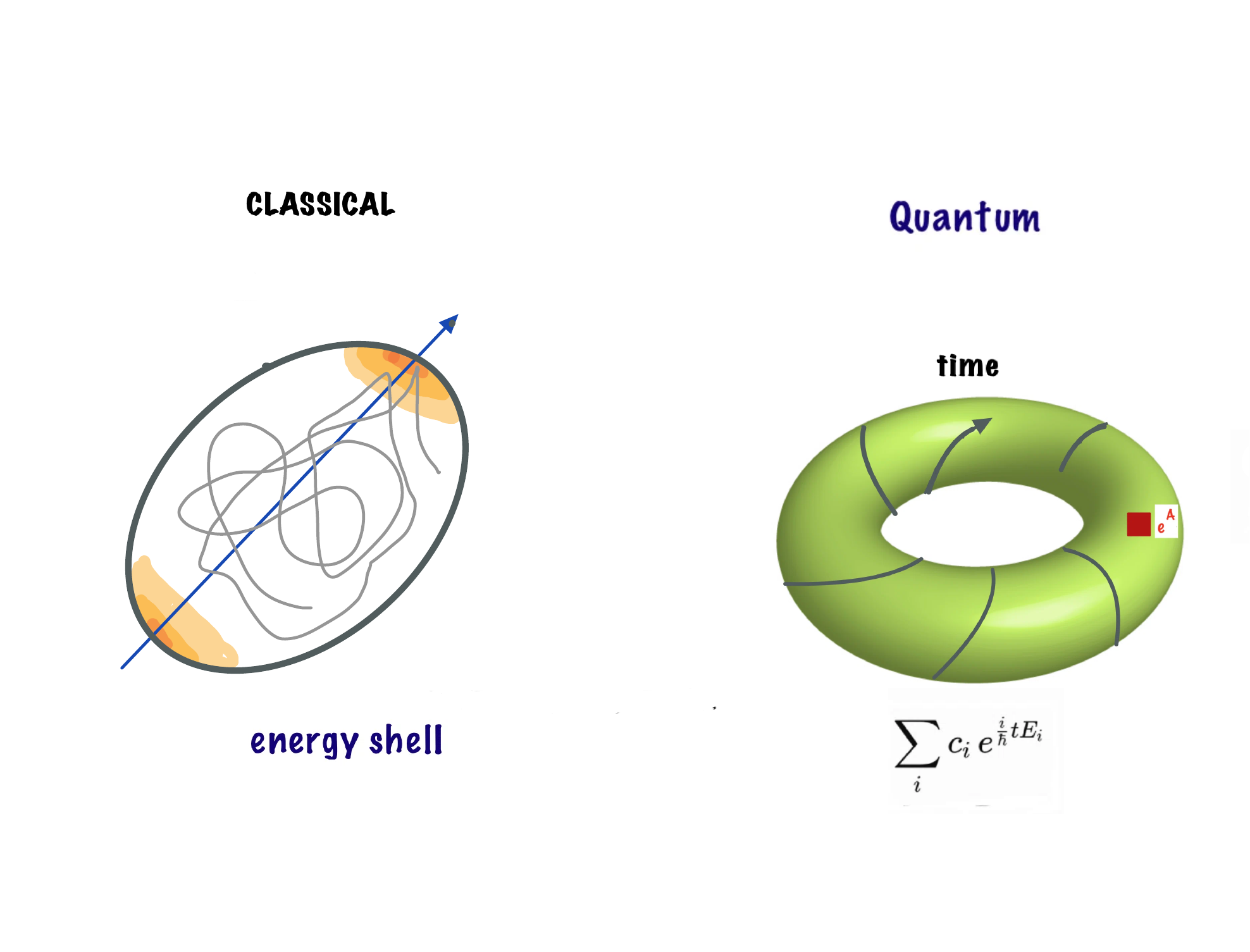}
    \caption{Large deviations. Left: classical on the energy shell, incursions on regions with rare values of $A$ are rare and brief. Right: same for expectation $\langle \Psi(t)|A|\Psi(t)\rangle$, motion is on the torus of phases.}
    \label{deviations}
\end{figure}

\paragraph{Classical}~\\
Consider a macroscopic  classical chaotic system evolving in the energy shell and a macroscopic observable $A(p,q)$. We may compute the flat probability distribution of $A$ on the shell, for example through $Z=e^{-\beta E -\beta^* A}$ adjusting $\beta$ to fix the desired energy and $\beta^*$ to obtain  through Legendre transform the probabilities in terms of $A$, which often takes the large deviation  form
$P(A)  = e^{NI(A/N)} $ - see Figure \ref{deviations}.  

\begin{figure}
    \centering
    \includegraphics[width=0.9
    \linewidth]{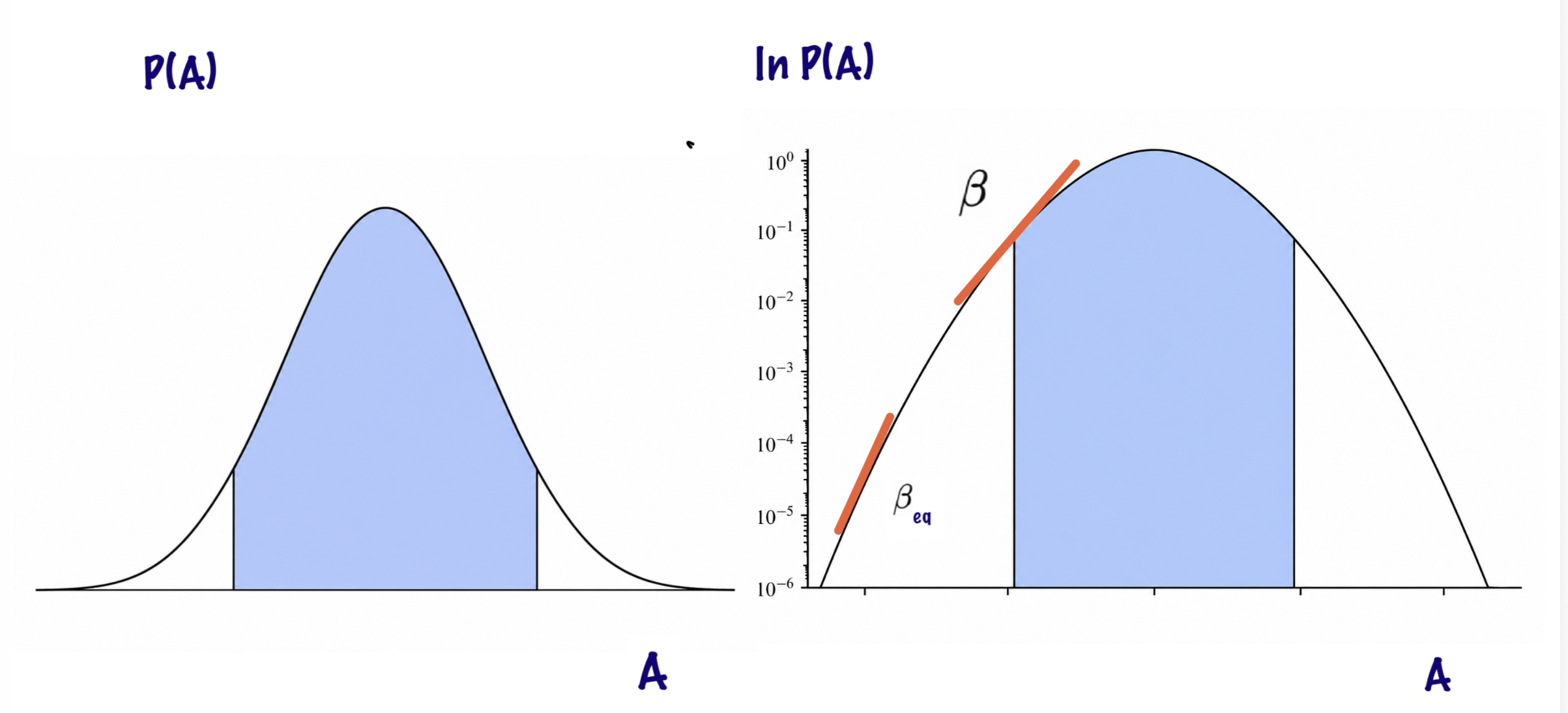}
    \caption{ The  crossover $\beta(t)$, above which  the boundary  $A^*(t)$ dominates, plays the role of a slowly time-dependent inverse temperature, which saturates at a value $\beta^*_{eq}$}
    \label{plots}
\end{figure}

\begin{figure}
    \centering
    \includegraphics[angle=-90,width=1.\linewidth]{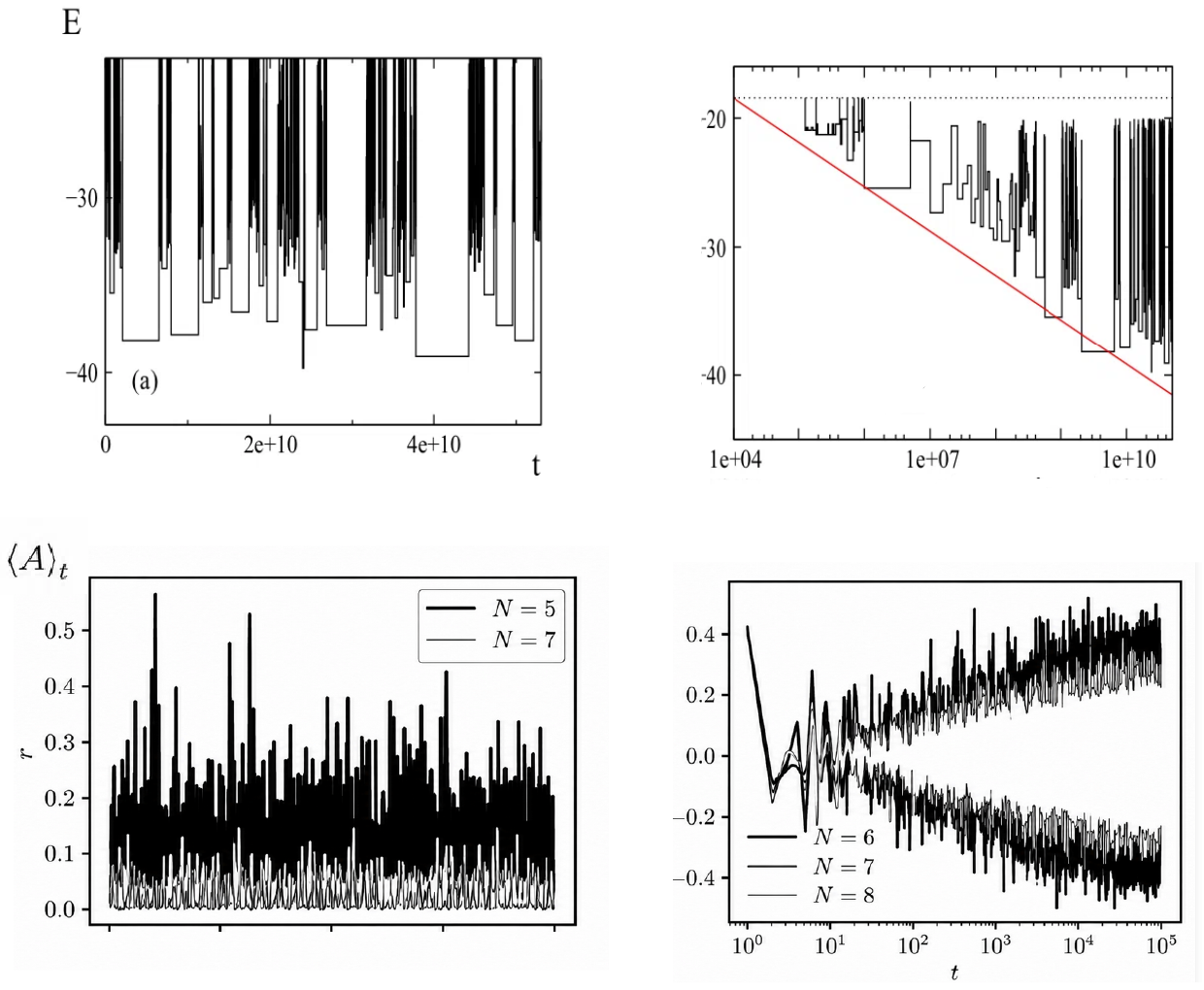}
    \caption{Time direction starting from a `typical'  configuration becomes evident when  plotted as a function of $\log t$. Above: energy in aging of a trap model, (from ref. \cite{junier2004microscopic}). Below: large deviations of a quantum spin system (from ref. \cite{CaoKurchan2026}). Left is linear, right logarithmic time.  }
    \label{fluctuations}
\end{figure}

Now consider the time evolution starting from a random configuration on the energy shell. 
As time progresses, the system makes deeper (brief) incursions into the zones of rare values of $A$,  at infinite times being able to resolve the full distribution of $P(A)$. At finite but long times, 
one obtains a distribution that looks like Figs \ref{plots} and \ref{fluctuations}: the thermodynamic $P(A)$ is reproduced, but with a cutoff at a time-dependent value $A^*(t)$, such that  the dynamics has not been able to go at time $t$  even once beyond $A^*(t)$. {\it The reason for this cutoff to be  sharp is the exponential dependence of $P(A)$:  just above $A^*$ the system never visited, but just below it did plenty of times.} The argument is familiar for the Random Energy Model \cite{derrida1981random} and from extreme value statistics \cite{bouchaud1997universality}. 

A direct consequence of this cutoff is that if we are computing the expectation of $e^{\beta_{eq} A}$ ($\beta_{eq}$ is our choice), we can do this by first computing the saddle 
point of $  \; e^{\ln P(A) +A {\beta^*_{eq}}} $, call it $A_{eq}$.  If $A_{eq}$ is inside the  interval visited so far, then the calculation is  correct.
If, instead, the saddle is outside that interval (the unshaded region in Fig. \ref{plots}), then the integral sticks to the boundary $A^*(t)$, and its corresponding $\beta^*$. 
All in all we have a situation where the dominant point of the integral sticks to the edge $A^*(t)$ of the explored region (with log derivative $\beta^*(t)$) until at some time $A^*(t_{eq})$ 
hits $A_{eq},\beta_{eq}$ and it freezes there.
{\it This is the `equilibration' time for the quantity $e^{\beta_{eq} A}$}
If we assume that the dynamics up to time $t$ samples of order $M \sim t/\tau$ independent points, where $\tau$ is a typical `forgetting'  time (expected to be at most polynomial in the system size), then 
$M \ln P(A^*(t)) \sim 1$ is an estimate for $A^*(t)$.

\paragraph{Quantum system} ~\\

In Ref. \cite{CaoKurchan2026} we have discussed the quantum counterpart of this. The more physical construction is to consider  dynamics monitored by weak measurements. Here, for simplicity, we discuss a simpler construction:  the evolution of 
${\cal A}(t)=\langle\Psi(t')|A |\Psi(t')\rangle$, and the associated probability distribution over $0<t'<t$ (see Fig \ref{deviations}). The argument is the same as the classical one above, except that the space being explored is the torus of quantum phases (Fig. \ref{deviations}) of the dimension of the Hilbert space ${\cal{N}} \sim e^{aN}$ .  The exploration is not chaotic, and the frequencies are correlated, but therein lies the interest.
Consider a quantum system with energy eigenvalues  $e_i$ and
an observable
\begin{equation}
{\cal A}(t)=\langle\Psi(t)|A |\Psi(t)\rangle =\sum_{ij}c_i c_j^* A_{ij} e^{i(e_i-e_j)t}.
\end{equation} Here $\langle i|\Psi_t\rangle=  c_i   e^{i e_i t}$.
Introducing phases $\theta_i=e_i t$, the evolution can be
represented as an integrable classical Hamiltonian system,
\begin{equation}
{\cal H}=\sum_i e_i I_i,
\qquad
\dot{\theta}_i=e_i,
\qquad
\dot I_i=0,
\end{equation}
where $I_i$ are conjugate to $\theta_i$. The observable
becomes
\begin{equation}
{\cal A}(\boldsymbol{\theta})=
\sum_{ij}c_i c_j^* A_{ij}e^{i(\theta_i-\theta_j)}.
\label{xym}
\end{equation}
(We use upper-case calligraphic letters ${\cal A, H, N, B}$ for the quantities that correspond to the statistical mechanics on the   torus  associated with Hilbert space phases, and $A,H,N,\beta$ for the original model.)
The `size' of the system being $\cal N$,  the large deviations are  $O( e^{{\cal N}})$, i.e. doubly exponential in $N$.

After Planckian times, the phases are independent, and the distribution is given by  flat averages over angles in the model with the XY spins Eq. (\ref{xym}).
The exploration is again the distribution  $P({\cal A})$ with a time-dependent cutoff just as in the classical case, see Fig.  \ref{fluctuations}. At shorter than Heisenberg times, the phases are not independent, but one can still argue that there will be a smooth  function of ${\cal A}$,  slowly varying with $t$, subjected to a  sharp cutoff as in the classical case separating frequently explored from unexplored \cite{CaoKurchan2026}.
{\it The cutoff value ${\cal A}^*(t)$ is then a measure of the complexity at time $t$ }. 
Estimates of the distribution near the cutoff can be made by studying the statistics of the most extreme values of ${\cal A}$ in the time-interval corresponding to independent spikes.

Let us note here that at "ramp" times of a fraction of Heisenberg time, the number of effective independent samplings is a deviation of ${\cal A}$ that is small with respect to double exponential, but  large with respect to typical $\sqrt {{\cal N}}$ fluctuations.\\

Two interesting choices for ${\cal A}$ are: \\

$\bullet$ The return amplitude $\langle\psi(0)|\psi(t)\rangle$ with
\begin{equation}
{ A} = |\psi_0\rangle\langle \psi_0|,
\end{equation}

$\bullet$ The spectral form factor:
 (without the log) as return amplitude of the thermo-field double:
\begin{equation}
|\Psi(0)\rangle=|\mathrm{TFD}(\beta)\rangle
=
\sum_n e^{-\beta E_n/2}
\,|n\rangle_L |n\rangle_R .
\end{equation}

\begin{equation}
|\mathrm{TFD}(\beta,t)\rangle
=
\sum_n
e^{-(\beta/2+it)E_n}
\,|n\rangle_L |n\rangle_R .
\end{equation}

\begin{equation}
{ A}= |\mathrm{TFD}(\beta)\rangle\langle \mathrm{TFD}(\beta)| \hspace{.4cm} \rightarrow \hspace{.4cm}
\langle \mathrm{TFD}(\beta,t)\rangle| A |\mathrm{TFD}(\beta,t)\rangle
=
\left|\sum_n e^{-(\beta+it)E_n}\right|^2
=
|Z(\beta+it)|^2.
\end{equation}
 
\paragraph{Reversibility between renewals }

Complexity has been conceived as a form of entropy but  one that continues to grow long after thermal equilibration. As such, it is expected to obey a kind of second law, exhibiting a persistent tendency to increase \cite{Susskind2020ThreeLectures}.
At this point, however, one may feel uneasy: in our system, long periods of aimless wandering through typical regions are punctuated by rare, large excursions of $A$. If given a sequence of values of $A(t)$,  we only look at it between two  "renewal" periods of typicality, we find that the sequence is {\it statistically time-reversible}. How does this match the idea of an arrow of time? Rather than attempting an explanation here, I will first describe how {\it exactly} the same situation arises in a model of an aging glass, in which it is clear that time has a preferred direction.

\section{Glassy dynamics and the Trap Model}

Glasses are by definition systems that lack an obvious form of order
and yet take extremely long - or perhaps forever - to reach equilibrium. 
This long-lasting  state of affairs, which is the one of all glasses in real life, is called `physical aging'. The two-time correlation functions 
have no time-translational invariance, they are of a form $C(t+t_w,t_w)$ and remain dependent on the waiting time $t_w$. Perhaps more surprisingly, they have {\it effective temperatures} that are usually defined by fluctuation-dissipation \cite{CugliandoloKurchanPeliti1997}.

Years ago, Bouchaud introduced a very minimal model of aging, the trap model \cite{bouchaud1995aging}. 
There are traps whose depths $E>0$ 
are exponentially distributed (see Fig. \ref{trap}),
\begin{equation}
\rho(E)=\frac{1}{T_g}e^{-E/T_g}.
\tag{1}
\end{equation}
The system  jumps from a trap by thermal activation,
returning to a common high energy threshold before falling into
a new trap chosen at random. The mean residence time in a
trap of depth $E$ is
\begin{equation}
\tau(E)=\tau_0 e^{E/T}.
\tag{2}
\end{equation}
Changing variables from trap depths to residence times gives
\begin{equation}
\rho(\tau)=\frac{T}{T_g}\tau_0^{T/T_g}
\tau^{-1-T/T_g},
\qquad \tau\geq\tau_0.
\tag{3}
\end{equation}
Thus an exponential distribution of energies becomes a
power-law distribution of trapping times, with
exponent $\mu=T/T_g$. For $T<T_g$, the mean waiting time
{\it diverges}, and the sum of many waiting times obeys a
L\'evy stable law of index $\mu<1$. This produces aging:
the longer the system has evolved, the longer it typically
takes to escape its current trap.
The correlation at $\mu=T/T_g<1$:
\begin{equation}
C(t_w+t,t_w)=
\frac{\sin(\pi\mu)}{\pi}
\int_{\frac{t}{t+t_w}}^1
\frac{du}{u^\mu(1-u)^{1-\mu}}.
\end{equation}
A  low-temperature realization is shown in figure \ref{fluctuations}.

\subsection*{A locally exponential tail: the running trap model}

\begin{figure}
    \centering
    \includegraphics[width=0.9\linewidth]{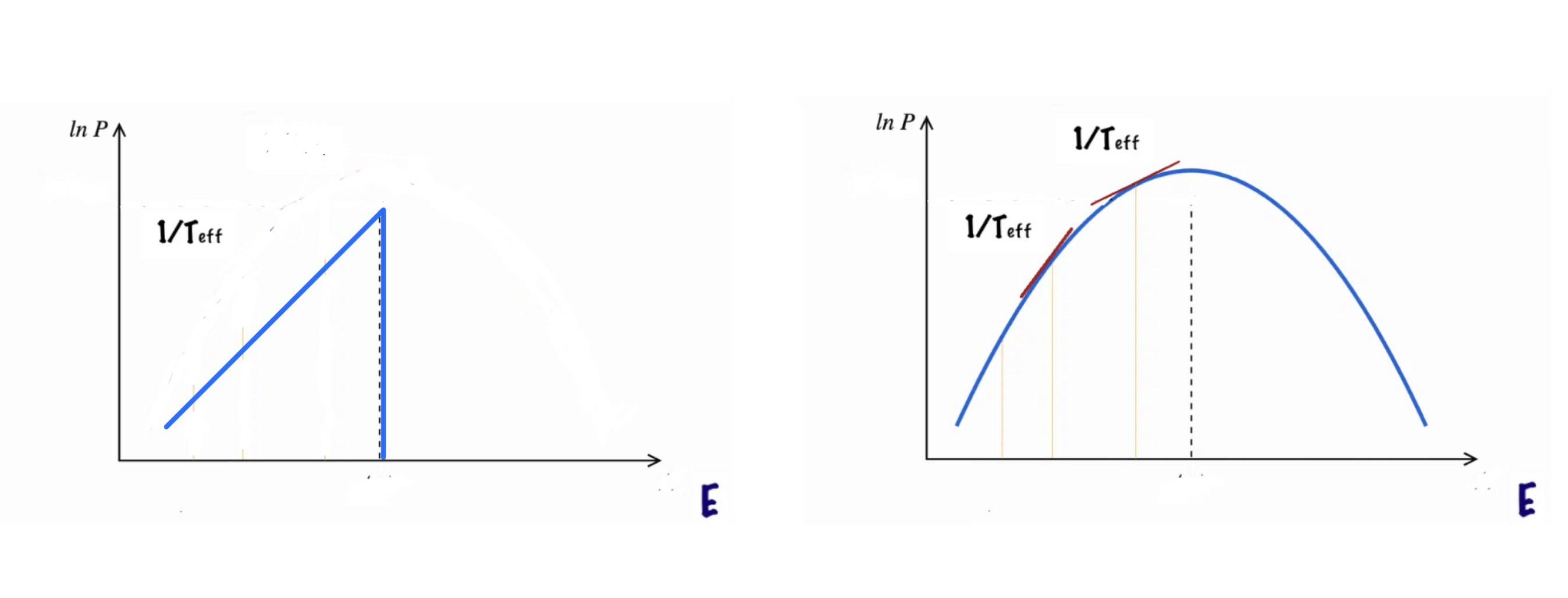}
    \caption{The trap model and its generalization. Right: the running effective temperature is the local derivative at each timescale, equilibration happens when it hits $T_{eq}$ }
    \label{trap}
\end{figure}

One may consider a more general trap model with a landscape with  exponentially many states $M$ and a distribution
of positive depths satisfying (Fig. \ref{trap})
\begin{equation}
p_N(E)\asymp \exp[-N I(E/N)],
\end{equation}
where $I(a)$ is an increasing 
function in the region of interest. 
At each extensive depth $E=aN$, the timescale for emerging is
of order $e^{\beta aN}$. Around a given `epoch', characterized by a, we may expand
 $E=Na+x$, with $x=O(1)$, and the distribution
is locally exponential:
\begin{equation}
\frac{p_N(Na+x)}{p_N(Na)}
\longrightarrow e^{-I'(a)x},
\qquad
\beta_g^{\rm eff}(a)={I'(a)}.
\end{equation}
This suggests a ``running'' trap model, with a scale-dependent effective glass temperature and aging exponent \footnote{This generalization of
Bouchaud's trap model was studied for the Random Energy
Model (REM) by Ben Arous, Bovier and Gayrard
\cite{BenArousBovierGayrard2002,
BenArousBovierGayrard2003I,BenArousBovierGayrard2003II}.
and further
investigated by Ben Arous and \v{C}ern\'y 
\cite{BenArousCerny2008}, and later extended to mean-field
spin glasses  \cite{BenArousBovierCerny2008}.}
Thermal activation gives residence times
$\tau\sim\tau_0 e^{\beta E}$. The corresponding local
power-law exponent is therefore
\begin{equation}
\alpha(a)=\frac{I'(a)}{\beta}.
\end{equation}
An observation time $t\sim\tau_0 e^{\gamma N}$
selects the depth $a=\gamma/\beta$, which yields
\begin{equation}
\alpha(\gamma)=\frac{1}{\beta}
I'\left(\frac{\gamma}{\beta}\right),
\qquad
\alpha(t)=\frac{1}{\beta}
I'\left(\frac{\log(t/\tau_0)}{\beta N}\right).
\end{equation}
Thus the effective exponent runs (very slowly) with observation
time. For a convex rate function, deeper traps
correspond to a steeper local exponential tail,
so that $T_g^{\rm eff}$ decreases and $\alpha$
increases.
In our construction,  the running
temperature $T(t)$ instead decreases until
\begin{equation}
T(t)=T_{\rm eq},
\qquad
\alpha(t)=1.
\end{equation}
see Fig. \ref{trap}. At this point the explored region reaches the equilibrium peak,
and aging crosses over to stationary dynamics. 

\paragraph{Reversibility between renewals (again)}
~\\~\\
If we consider the energies visited by a trap model between renewals, i.e.  returns to the high energy threshold, the trajectories are statistically symmetrical under time reversal. And yet, the system ages.  
 One has to bear in mind here that any system
having detailed balance has statistical  time-reversal between 
instants with the same energy, and yet when starting from an unlikely initial condition there is a time-arrow towards equilibration (followed by a recurrence  much later). In our case, equilibration time is infinite in the original trap model, and it is the time it takes to reach ${\beta}_{eq}$ in the more general one.

 In the quantum case, we will see that the analogous statement is that the arrow comes from starting at $t=0$ with a specific $|\Psi(0)⟩$, not from any irreversibility of the quantum dynamics itself. This distinction is physically important and worth making explicit.

\section{The mapping}

\subsection*{Relation between complexity and trap model for a classical evolution}

Consider a classical system with N degrees of freedom,
whose energy surface is explored by an ergodic dynamics.
Let A be an extensive observable, A=Na, satisfying
a large-deviation principle
\begin{equation}
P_N(A=Na)\asymp e^{-NI(a)}.
\end{equation}
Let us rescale times as:
\begin{equation}
\frac{dt}{dt'}=e^{-\beta_{eq} A(t)},
\qquad
dt'=e^{\beta_{eq} A(t)}dt.
\end{equation}
The measure over time $t'$ is now biased by $e^{\beta_{eq} A}$, rare regions are `traps' where dynamics is extra slow.
Assuming sufficiently rapid mixing in the original time,
successive visits to rare regions may be regarded as
approximately independent. Their statistics is governed
by the large-deviation function I(a), while their
residence times in the rescaled dynamics grow as
\begin{equation}
\tau(a)\sim\tau_0 e^{\beta Na}.
\end{equation}
The rescaled dynamics therefore explores the same
phase-space regions as a trap model whose trap depths
are Na and whose density is governed by the same
large-deviation function I(a). The equivalence concerns
the statistics of trapping and exploration.

\subsection*{Quantum phase dynamics as a reversible trap model}

Let us see now how this applies to the large deviations of a quantum expectation. For simplicity, we concentrate on the spectral form factor, the return probability of a thermo-field double, at infinite temperatures.
Unlike a generic observable, whose
expectation may have approximately Gaussian fluctuations,
the SFF has an approximately exponential distribution.
The statistics of its low-order moments have been studied
by Altland et al.~\cite{Altland2025} and subsequently by
Charamis et al.~\cite{Charamis2026}. These results motivate
an exponential distribution with a time-dependent mean,
crossing over from the ramp to the plateau.
Writing
\begin{equation}
a=\mathcal A(\theta)
=\frac{1}{{\cal N}^2}
\left|\sum_{j=1}^{{\cal N}}e^{i\theta_j}\right|^2,
\end{equation}
then:
\begin{equation}
P_t(a)\simeq
\frac{1}{\bar a(t)}
\exp\left[-\frac{a}{\bar a(t)}\right].
\end{equation}
In the unfolded unitary random-matrix setting,
up to convention-dependent factors,
\begin{equation}
\bar a(t)\sim
\begin{cases}
t/{\cal N}^2, & t\ll{\cal N}
\quad\text{(ramp)},\\
1/{\cal N}, & t\gtrsim{\cal N}
\quad\text{(plateau)}.
\end{cases}
\end{equation}
We assume that, beyond the scrambling time, sufficiently
separated excursions can be treated as independent, with
an effective sampling interval $\tau$. We also assume
that the exponential form remains valid up to the
moderate-deviation scale (i.e. exponential rather than doubly exponential) explored by these samples.

The distribution sampled along a trajectory up to time
$t$ is then approximately smooth below a moving
extreme-value cutoff determined by
\begin{equation}
\frac{1}{\tau}\int_0^t ds\,
\exp\left[-\frac{a_c(t)}{\bar a(s)}\right]
\sim 1.
\end{equation}
Since $\bar a(s)$ increases during the ramp with $s$, this gives, to
leading logarithmic accuracy,
\begin{equation}
a_c(t)\simeq
\bar a(t)\ln\frac{t}{\tau}.
\end{equation}
and the same expression holds on the plateau, where
$\bar a(t)$ becomes time independent. Thus, in both
regimes, the explored distribution is approximately
exponential, truncated at a cutoff that moves towards
larger values as sampling proceeds.

In order to map this to a trap model, we  introduce the reparametrized clock
\begin{equation}
\frac{dt}{d\tau'}
=e^{-{\cal B}_{eq}a},
\end{equation}
 Large excursions of $a$ are
thereby converted into long trapping episodes, with
characteristic durations
\begin{equation}
\tau_{\mathrm{trap}}(a)
\sim e^{{\cal B}_{eq}a}.
\end{equation}
The exponential distribution of $a$ becomes a power-law
distribution of trapping times,
\begin{equation}
\mathrm{Prob}(\tau_{\mathrm{trap}}>u)
\sim u^{-\mu(t)},
\qquad
\mu(t)=\frac{1}{{\cal B}_{eq}\bar a(t)}.
\end{equation}
To make trapping relevant on timescales of order
${\cal N}$, we choose ${\cal B}_{eq} \propto {\cal N}$:
\begin{equation}
{\cal B}_{eq}=b{\cal N}.
\end{equation}
The exponent then decreases during the ramp and reaches
the constant plateau value
\begin{equation}
\mu=\frac{1}{b}.
\end{equation}
For $b>1$, this lies in the aging regime $\mu<1$ of
the Bouchaud trap model.

At the moving cutoff, the largest trapping times are
\begin{equation}
\tau_{\mathrm{trap}}(a_c(t))
\sim
\left(\frac{t}{\tau}\right)^{
{\cal B}_{eq}\bar a(t)}.
\end{equation}
In particular, at the Heisenberg time
$t\sim{\cal N}$,
\begin{equation}
a_c({\cal N})\sim
\frac{\ln({\cal N}/\tau)}{{\cal N}},
\qquad
\tau_{\mathrm{trap}}(a_c({\cal N}))
\sim({\cal N}/\tau)^b.
\end{equation}
Since ${\cal N}\sim e^{sN}$, the resulting trapping
times are exponential rather than doubly exponential
in $N$.

Under the assumed independence of rare excursions,
 the time-reparametrized dynamics
 again maps onto a trap model.
The exponent runs during  ramp times and becomes constant
in Heisenberg times, where the standard Bouchaud aging
regime is recovered for $b>1$.

Finally, the time reparametrization changes neither the
trajectories nor their invertibility. When the original
dynamics has a time-reversal symmetry under which
$a$ is even, this symmetry is preserved. We thus obtain
a deterministic, reversible realization of trap-like aging:
the traps are not metastable minima, but rare configurations
of quantum phases, whose statistics is governed by the
large-deviation function $I(a)$.

\section{Conclusion}

The picture that emerges is one in which  extensive observables, or their quantum expectations, are being sampled in the (so defined) equilibration time. 
`Rare' quantities, such as the exponent  of a measurement or of a quantum expectation (or partial measurement \cite{CaoKurchan2026}), take their final value each at a sharply defined, much longer time, a kind of `second equilibration'.

Once mapped onto a model of glassy dynamics, this becomes a situation of long-time aging of the observable, finally reaching equilibrium.
The second law becomes the fact that the number of effective independent samplings $M(t)$ performed by the dynamics increases linearly with overwhelming probability over time, although  a very unusual trajectory  could in principle abstain from going into rare places for long times, or may go there often.

For aging in general glass models there is a variety of situations we know: many timescales and many temperatures,  memory effects, correlations between states. 
It remains to be seen whether this richness may reflect in the physics of expectation of rare observables at post-equilibration times: it would arise if the space of deviations had a non trivial structure.

~\\~\\

{\bf \large Acknowledgments}\\~\\
This work is dedicated to Uzi Smilansky on occasion of his birthday. I am grateful to Xiangyu Cao for our earlier collaboration, on which much of the present work builds.

\bibliographystyle{unsrt}
\bibliography{paper.bib}

@article{uogh,
	title = {A Universal Operator Growth Hypothesis},
	author = {Parker, Daniel E. and Cao, Xiangyu and Avdoshkin, Alexander and Scaffidi, Thomas and Altman, Ehud},
	journal = {Phys. Rev. X},
	volume = {9},
	issue = {4},
	pages = {041017},
	numpages = {29},
	year = {2019},
	month = {Oct},
	publisher = {American Physical Society},
	doi = {10.1103/PhysRevX.9.041017},
	url = {https://link.aps.org/doi/10.1103/PhysRevX.9.041017}
}

@article{Altland2025,
  author  = {Altland, Alexander and Divi, Felix and
             Micklitz, Tobias and Pappalardi, Silvia and
             Rezaei, Mohammad},
  title   = {Statistics of the Random Matrix Spectral Form Factor},
  journal = {Physical Review Research},
  volume  = {7},
  pages   = {033138},
  year    = {2025},
  doi     = {10.1103/PhysRevResearch.7.033138}
}

@article{Charamis2026,
  author  = {Charamis, Dimitrios and Kulkarni, Manas and
             Kurchan, Jorge and Foini, Laura},
  title   = {Quenched Properties of the Spectral Form Factor},
  journal = {Physical Review E},
  volume  = {113},
  pages   = {014138},
  year    = {2026},
  doi     = {10.1103/PhysRevE.113.014138}
}

@article{nielsen06-1,
    author = {Nielsen, Michael A.},
    title = {A geometric approach to quantum circuit lower bounds},
    journal =  {Quantum Inf. Comput.},
    year = {2006},
    volume = {6},
    pages= {213},
    doi = {10.26421/QIC6.3-2}
}

@article{nielsen06,
author = {Michael A. Nielsen and Mark R. Dowling and Mile Gu  and Andrew C. Doherty },
title = {Quantum Computation as Geometry},
journal = {Science},
volume = {311},
number = {5764},
pages = {1133-1135},
year = {2006},
doi = {10.1126/science.1121541},
URL = {https://www.science.org/doi/abs/10.1126/science.1121541},
eprint = {https://www.science.org/doi/pdf/10.1126/science.1121541}}

@article{Rabinovici:2020ryf,
    author = "Rabinovici, E. and S{\'a}nchez-Garrido, A. and Shir, R. and Sonner, J.",
    title = "{Operator complexity: a journey to the edge of Krylov space}",
    eprint = "2009.01862",
    archivePrefix = "arXiv",
    primaryClass = "hep-th",
    doi = "10.1007/JHEP06(2021)062",
    journal = "JHEP",
    volume = "06",
    pages = "062",
    year = "2021"
}

@misc{Rabinovici:2025otw,
    author = "Rabinovici, Eliezer and S{\'a}nchez-Garrido, Adri{\'a}n and Shir, Ruth and Sonner, Julian",
    title = "{Krylov Complexity}",
    eprint = "2507.06286",
    archivePrefix = "arXiv",
    primaryClass = "hep-th",
    reportNumber = "CERN-TH-2025-128",
    month = "7",
    year = "2025"
}

@article{vijay,
	title = {Quantum chaos and the complexity of spread of states},
	author = {Balasubramanian, Vijay and Caputa, Pawel and Magan, Javier M. and Wu, Qingyue},
	journal = {Phys. Rev. D},
	volume = {106},
	issue = {4},
	pages = {046007},
	numpages = {28},
	year = {2022},
	month = {Aug},
	publisher = {American Physical Society},
	doi = {10.1103/PhysRevD.106.046007},
	url = {https://link.aps.org/doi/10.1103/PhysRevD.106.046007}
}

@article{nandy-review,
	title = {Quantum dynamics in Krylov space: Methods and applications},
	journal = {Physics Reports},
	volume = {1125-1128},
	pages = {1-82},
	year = {2025},
	note = {Quantum dynamics in Krylov space: Methods and applications},
	issn = {0370-1573},
	doi = {https://doi.org/10.1016/j.physrep.2025.05.001},
	url = {https://www.sciencedirect.com/science/article/pii/S0370157325001462},
	author = {Pratik Nandy and Apollonas S. Matsoukas-Roubeas and Pablo Martínez-Azcona and Anatoly Dymarsky and Adolfo {del Campo}}
}

@article{krylov-nielsen,
  title = {A Relation between Krylov and Nielsen Complexity},
  author = {Craps, Ben and Evnin, Oleg and Pascuzzi, Gabriele},
  journal = {Phys. Rev. Lett.},
  volume = {132},
  issue = {16},
  pages = {160402},
  numpages = {6},
  year = {2024},
  month = {Apr},
  publisher = {American Physical Society},
  doi = {10.1103/PhysRevLett.132.160402},
  url = {https://link.aps.org/doi/10.1103/PhysRevLett.132.160402}
}

@article{derrida1981random,
  title = {Random-energy model: An exactly solvable model of disordered systems},
  author = {Derrida, Bernard},
  journal = {Phys. Rev. B},
  volume = {24},
  issue = {5},
  pages = {2613--2626},
  numpages = {0},
  year = {1981},
  month = {Sep},
  publisher = {American Physical Society},
  doi = {10.1103/PhysRevB.24.2613},
  url = {https://link.aps.org/doi/10.1103/PhysRevB.24.2613}
}

@article{bouchaud1997universality,
 doi = {10.1088/0305-4470/30/23/004},
url = {https://doi.org/10.1088/0305-4470/30/23/004},
year = {1997},
month = {dec},
publisher = {},
volume = {30},
number = {23},
pages = {7997},
author = {Jean-Philippe Bouchaud and Marc Mézard},
title = {Universality classes for extreme-value statistics},
journal = {Journal of Physics A: Mathematical and General}
}

@misc{CaoKurchan2026,
  author        = {Cao, Xiangyu and Kurchan, Jorge},
  title         = {Large deviations in quantum dynamics and complexity},
  year          = {2026},
  eprint        = {2607.20959},
  archivePrefix = {arXiv},
  primaryClass  = {cond-mat.stat-mech},
  url           = {https://arxiv.org/abs/2607.20959}
}

@incollection{Susskind2020ThreeLectures,
  author = {Susskind, Leonard},
  title = {Three Lectures on Complexity and Black Holes},
  booktitle = {Quantum Information and Quantum Gravity},
  publisher = {Springer},
  year = {2020},
  pages = {175--217},
  doi = {10.1007/978-3-030-45109-7_5}
}

@article{Bouchaud1992,
  author = {Bouchaud, Jean-Philippe},
  title = {Weak ergodicity breaking and aging in disordered systems},
  journal = {Journal de Physique I}, volume = {2}, pages = {1705--1713}, year = {1992}
}

@article{junier2004microscopic,
  title={Microscopic realizations of the trap model},
  author={Junier, Ivan and Kurchan, Jorge},
  journal={Journal of Physics A: Mathematical and General},
  volume={37},
  number={13},
  pages={3945--3965},
  year={2004}
}

@article{bouchaud1995aging,
  title={Aging on Parisi's tree},
  author={Bouchaud, J-P and Dean, David S},
  journal={Journal de Physique I},
  volume={5},
  number={3},
  pages={265--286},
  year={1995}
}

@article{CugliandoloKurchanPeliti1997,
  author  = {L. F. Cugliandolo and J. Kurchan and L. Peliti},
  title   = {Energy flow, partial equilibration, and effective
             temperatures in systems with slow dynamics},
  journal = {Physical Review E},
  volume  = {55},
  pages   = {3898--3914},
  year    = {1997},
  doi     = {10.1103/PhysRevE.55.3898},
  eprint  = {cond-mat/9611044},
  archivePrefix = {arXiv}
}

@article{BenArousBovierGayrard2002,
  author = {Ben Arous, G. and Bovier, A. and Gayrard, V.},
  title = {Aging in the Random Energy Model},
  journal = {Physical Review Letters},
  volume = {88},
  pages = {087201},
  year = {2002},
  doi = {10.1103/PhysRevLett.88.087201}
}

@article{BenArousBovierGayrard2003I,
  author = {Ben Arous, G. and Bovier, A. and Gayrard, V.},
  title = {Glauber Dynamics of the Random Energy Model.
           I. Metastable Motion on the Extreme States},
  journal = {Communications in Mathematical Physics},
  volume = {235},
  pages = {379--425},
  year = {2003}
}

@article{BenArousBovierGayrard2003II,
  author = {Ben Arous, G. and Bovier, A. and Gayrard, V.},
  title = {Glauber Dynamics of the Random Energy Model.
           II. Aging Below the Critical Temperature},
  journal = {Communications in Mathematical Physics},
  volume = {236},
  pages = {1--54},
  year = {2003},
  doi = {10.1007/s00220-003-0799-3}
}

@article{BenArousCerny2008,
  author = {Ben Arous, G. and {\v C}ern{\'y}, J.},
  title = {The Arcsine Law as a Universal Aging Scheme
           for Trap Models},
  journal = {Communications on Pure and Applied Mathematics},
  volume = {61},
  pages = {289--329},
  year = {2008},
  doi = {10.1002/cpa.20177}
}

@article{BenArousBovierCerny2008,
  author = {Ben Arous, G. and Bovier, A. and {\v C}ern{\'y}, J.},
  title = {Universality of the {REM} for Dynamics
           of Mean-Field Spin Glasses},
  journal = {Communications in Mathematical Physics},
  volume = {282},
  pages = {633--695},
  year = {2008},
  doi = {10.1007/s00220-008-0565-7}
}

\end{document}